\documentclass[english]{revtex4-1}
\usepackage[T1]{fontenc}
\usepackage[latin9]{luainputenc}
\usepackage{textcomp}
\usepackage{amsmath}
\usepackage{amssymb}
\usepackage{graphicx}
\usepackage{babel}
\begin{document}

\title{Gravitational Quantum Well as an Effective Quantum Heat Engine }

\author{Jonas F. G. Santos }

\affiliation{Federal University of ABC - Santo André - São Paulo - Brazil}
\email{jonas.floriano@ufabc.edu.br}

\begin{abstract}
In this work the gravitational quantum well is used to model an effective
two level system and to perform two thermodynamic cycles, the isogravitational
and the isoenergetic ones. It is shown that the isogravitational
is independent of the scale parameter whereas the isoenergetic has
a dependence on the eigenstates chosen to form the cycle. An equivalent equation for the isoenergetic cycle 
is also obtained, which is similar to the equation of state for an isothermal process of an ideal
gas. This equation reinforces the concept of energy bath, where
the temperature is replaced by the energy into the expression of efficiency.
\end{abstract}
\maketitle

\section{Introduction}

Quantum heat engines are devices which transform an amount of heat
(thermal energy) into work (mechanical energy) in a scale where the
quantum effects are relevant and can be useful in some way to improve
this conversion of energy \cite{Gold01,Nori,Sanders}. In these cases,
the working substance is a quantum system such as a single \cite{Rezek01}
or many \cite{Reid01} harmonic oscillators, single-atom \cite{Singer},
single-ion \cite{Lutz02}, vacuum forces through the Casimir interaction
\cite{Omar}, quantum rotors \cite{Stella}, or a two level system (TLS) normally characterized as
the spin of some molecule \cite{Altintas,Batalhao} or atom. For the
cases where the thermal reservoirs are considered classical, the efficiency
of these engines surprisingly does not surpasses the classical limit
of Carnot, $\eta^{Carnot}=1-T_{c}/T_{h}$, where $T_{c}$ and $T_{h}$
are the temperatures of the cold and hot reservoirs, respectively
\cite{Gold01}. The same bound is no longer applicable when the reservoirs
are quantum in some level such as, for instance, when the working
substance interacts with a squeezed thermal reservoir \cite{Lutz,Togan},
when there is a protocol to extract the so-called imperfect work \cite{Mischa, NG},
or in the case of a heat engine based on off-resonant light interaction
\cite{Ritsch}, expliciting the role of quantum effects in the conversion
of energy.

Another possibility of changing the reservoirs such that we move from
a classical to a quantum perspective was introduced by Bender \cite{Bender01}.
He considered the concept of an energy reservoir in substitution of
a thermal reservoir in the quantum heat engine. Thermal reservoirs
are characterized by a well defined temperature associated with a stroke
where the working substance interacts exchanging heat. On the other
hand, the energy bath is such that during the correspondent stroke
the expectation value of the Hamiltonian, $\left\langle \hat{H}\right\rangle $,
of the working substance is kept fixed \cite{Pena02}. Such a stroke
is called isoenergetic. The applicability of isoenergetic strokes
has been extended to nonrelativistic regimes such as in the case
of a single particle confined into a cylindrical potential and submitted
to an external magnetic field \cite{Pena01}, to the noncommutative
version of quantum mechanics \cite{Santos01}, in the relativistic
regime of a single-particle Dirac spectrum \cite{Pena03}, and for
the Habi model \cite{Pena02}.

On the basis of two level systems, the isoenergetic cycles can be
effectively modeled by considering the two first states of some particular
system provided one has sufficient control in order to avoid
that others states are occupied. Once this condition is fulfilled,
thermodynamic cycles can be performed. In refs.\cite{Bender01,Ou},
the authors considered an isoenergetic stroke where the length of
an infinity square well is quasi-statically changed from $L$ to $L+\Delta L$.
The relevant point here is the existence of a length scale that can
be changed by using the variation of some external agent in an isoenergetic
stroke.

Based on the argument above, the gravitational quantum well (GQW)
is a suitable system where the isoenergetic stroke can be tested
by using the intensity of the gravitational interaction as an external
field. The GQW system is important because it is possible
to obtain bound states for a particle coupled to the gravitational
field. In ref. \cite{Neutron01}, the spacial distribution of ultracold
neutrons coupled to Earth by gravitational interaction was experimentally
measured and it is consistent with the theoretical result using
the Wigner function. Moreover, there are several recent studies employing
the GQW system as a base of test for generalizations of quantum theory,
for instance, in the case of a deformed Heisenberg algebra \cite{Brau01,Bertolami,Gouba}. The basic idea is to use the GQW architecture as a fuel to a quantum heat engine by considering the two first eigenstates, given by the Airy functions.
Introducing a gravitational length scale, $\ell_{g}=(\hbar^{2}/2m^{2}g)^{1/3}$, two different thermodynamic cycles based on the GQW system are assumed.
The former one is called isogravitational, composed by two isogravitational and
two isoentropic strokes, and the latter one by
two isoenergetic and two isoentropic strokes and called isoenergetic cycle. From an experimental
perspective, recent development in nanoscale experimental techniques
made it possible to engineer realistic quantum heat engines \cite{Casati},
for instance, using a single molecule junctions \cite{Chen}, single-level
quantum dot \cite{Markus}, optomechanical systems \cite{mari} etc.
Such a nanoscale quantum heat engines have demonstrate the fine control
in order to justify the experimental realization of an isoenergetic
cycle.

This work is organized as follows. In the next section we review the
eigenvalue equation for the Hamiltonian of a particle coupled to a
linear potential and then particularize for the GQW system. Section
\ref{sec:The-GQW-as} is devoted to describe in detail the two cycles
and discuss the analytical expressions for the efficiencies. We generalize
the possibility of generating a two level system with any pair of
eigenstates of the GQW system in section \ref{sec:general-iso-energetic-cycle}.
In section \ref{sec:Equation-of-State} we extend our analyzes in
order to obtain an analogous of equation of state for an isoenergetic
stroke and compare it with the same expression for an isothermal process
for an ideal gas. The conclusions and final considerations are drawn
in section \ref{sec:Conclusions}.

\section{Review on the Quantum Mechanics for Linear Potential\label{sec:Review-on-the}}

Here we will review the important aspects of the system composed by
a particle of mass $m$ into a linear potential and then restrict
the solutions to the particular case of the gravitational quantum
well. Let one consider a quantum system described by the one-dimensional Hamiltonian,
\begin{equation}
\mathbf{H}(x,\mathbf{p})=\frac{\mathbf{p}^{2}}{2m}+C\,x,
\end{equation}
where $C$ is an arbitrary constant such that the product $Cx$ has
dimension of energy. The eigenstates and eigenenergies are obtained
by solving the eigenvalue equation,
\begin{align}
\mathbf{H}\,\psi_{n} & =E_{n}\,\psi_{n}\\
\frac{d^{2}\psi}{dx^{2}}+\frac{2m}{\hbar^{2}} & (E_{n}-C\,x)\psi_{n}=0.\label{SH01}
\end{align}

Defining a new variable, 
\[
\xi=\left(x-\frac{E_{n}}{C}\right)\left(\frac{2mC}{\hbar^{2}}\right)^{1/3},
\]
eq. (\ref{SH01}) can be written in the form,
\begin{equation}
\frac{d^{2}\psi_{n}}{d\xi^{2}}+\xi\psi_{n}=0,\label{SH02}
\end{equation}
which is the Airy equation whose solutions are called Airy functions.
The general form for these solutions are \cite{Bookairy},
\begin{equation}
\psi(\xi)=D_{1}A_{i}(-\xi)+D_{2}B_{i}(-\xi),
\end{equation}
but as $B_{i}(-\xi)$ diverges for $x>0$, the physical meaning of
$\psi(\xi)$ imposes that $D_{2}=0$. The constant $D_{1}$ is found
by normalizing the wave function,
\begin{equation}
\int_{-\infty}^{\infty}\psi(\xi)\psi^{\ast}(\xi\text{\textasciiacute})d\xi=\delta(E-E\text{\textasciiacute}),
\end{equation}
resulting in $D_{1}=(2m)^{1/3}/\hbar^{2/3}C^{1/6},$ with the following eigenstates,
\begin{equation}
\psi_{n}(x)=\frac{(2m)^{1/3}}{\hbar^{2/3}C^{1/6}}A_{i}\left[-\frac{(2mC)^{1/3}}{\hbar^{2/3}}\left(x-\frac{E_{n}}{C}\right)\right].\label{eigenfu01}
\end{equation}

In the present case, we are interested in the gravitational quantum well (GQW),
i. e. when $C=mg$. Thus,
\begin{equation}
\psi_{n}(x)=\frac{(2m)^{1/3}}{\hbar^{2/3}mg^{1/6}}A_{i}\left[-\left(\frac{2m^{2}g}{\hbar^{2}}\right)^{1/3}\left(x-\frac{E_{n}}{mg}\right)\right],\label{eigenfu03}
\end{equation}
where the eigenenergies are obtained by imposing that $\psi_{n}(x=0)=0,$
which results in,
\begin{equation}
E_{n}=-\left(\frac{mg^{2}\hbar^{2}}{2}\right)^{1/3}a_{n},\label{Ener01}
\end{equation}
where $a_{n}$ are the zeroes of the Airy functions.

The GQW architecture is important because it possesses bound states
due to the gravitational coupling and has been tested in laboratory using ultracold neutrons (UCN) \cite{Neutron01}, where the spatial
distribution was experimentally obtained and it agrees with the theoretical
results via Wigner function of the system.

\section{The GQW as an effective two level system\label{sec:The-GQW-as}}

Following the framework introduced in the previous section, the manipulation of the GQW system as an effective two level system can now be investigated. This was originally done in ref. \cite{Pena01}, where a particle confined into a cylindrical potential and under the action of an external magnetic field was considered. In order to clarify
the physical meaning of the effective two level system and the strokes
involved, it is convenient to rewrite the eigenenergies (\ref{Ener01}) as,
\begin{equation}
E_{n}(g)=-\hbar\Omega_{g_{0}}\left(\frac{g}{g_{0}}\right)^{2/3}\alpha_{n},\label{Ener02}
\end{equation}
where $\Omega_{g_{0}}\equiv(mg_{0}^{2}/2\hbar)^{1/3}$ has unity of
frequency and $(g/g_{0})^{2/3}$is a dimensionless quantity. Thus
the energy is dependent explicitly on the intensity of the gravitational
field.

We are interested in two different cycles as illustrated in fig.\ref{Cycles}.
The former one is the isoenergetic cycle and was originally proposed by
Bender \cite{Bender01}, having been studied in different contexts
\cite{Pena01,Santos01,Pena02}. It is composed by two isoentropic
and two isoenergetic strokes. The isoenergetic process is performed
theoretically replacing the thermal bath model by an energy bath one
\cite{Pena02} and has the property that during the process the expectation
value of the Hamiltonian is kept fixed. The latter one will be called
here of isogravitational cycle whose is composed of two isoentropic and
two isogravitational strokes. A similar type of cycle was performed
in ref. \cite{Santos01} in the case of an external magnetic field. The
isogravitational stroke is a new one introduced theoretically here,
where the gravitational field is kept fixed during the stroke, with
the system performing a transition from the energy $E_{n}(g_{1})\rightarrow E_{m}(g_{1}).$
We will start our analysis studying the isogravitational cycle because
it is mathematically simpler.

\subsection{The Isogravitational Cycle}

This cycle is composed of two isoentropic and two isogravitational
strokes as depicted in fig.\ref{Cycles}. The isogravitational stroke is exactly as the isoenergy gap process of ref. \cite{Beretta}, where the associated frequency is dependent on the gravitational interaction. The system starts at the
ground state $\psi_{1}(g_{1})$ with energy $E_{1}(g_{1})$ and is
quasi-statically moved to the first excited state $\psi_{2}(g_{1})$
with energy $E_{2}(g_{1}).$ As in this stroke the gravitational field,
the only external agent, is kept fixed, the work performed on the
system is zero and the change in energy is exclusively called heat
and given by,
\begin{eqnarray}
\left\langle Q\right\rangle _{1\rightarrow2} & = & E_{2}(g_{1})-E_{1}(g_{1})\nonumber \\
 & = & -\hbar\Omega_{g_{0}}\left(\frac{g_{1}}{g_{0}}\right)^{2/3}(a_{1}-a_{2})<0.\label{Q12}
\end{eqnarray}

The second stroke is an isoentropic expansion and there is no heat
exchange. However, the system is driven from $\ell_{g_{1}}$to $\ell_{g_{2}}$and
then we can define an expansion coefficient $\alpha\equiv\ell_{g_{2}}/\ell_{g_{1}}$.
The third stroke is an isogravitational one such that it moves the
system from the state $\psi_{2}(g_{2})$ to $\psi_{1}(g_{2})$ with
$g_{2}<g_{1}.$ The heat exchange is then given by,
\begin{eqnarray}
\left\langle Q\right\rangle _{3\rightarrow4} & = & E_{1}(g_{2})-E_{2}(g_{1})\nonumber \\
 & = & -\hbar\Omega_{g_{0}}\left(\frac{g_{1}}{g_{0}}\right)^{2/3}\frac{1}{\alpha^{2}}(a_{2}-a_{1})>0,\label{Q34}
\end{eqnarray}
where the definition of $\alpha$ was considered The last stroke is
an isoentropic compression and again there is no heat exchange. By
observing the convention of signal of heat exchange, i. e. it is
positive when absorbed and negative when released by the system, we
can define the thermodynamic efficiency as,
\begin{equation}
\eta=\frac{\left\langle W\right\rangle}{\left\langle Q\right\rangle_{3\rightarrow4}}=1-\frac{\left\langle Q\right\rangle_{1\rightarrow2}}{\left\langle Q\right\rangle_{3\rightarrow4}}=1-\frac{1}{\alpha^{2}},\label{effIG}
\end{equation}
i. e. the efficiency of an isogravitational cycle does not depend
on the particular choice of $g_{0}$ and becomes close to one when
$\alpha$ is very large, which physically means an extremely difference
between $g_{1}$ and $g_{2}$. Thus, in a realistic point of view,
the efficiency of this type of cycle will be very small in practice.
The result in eq. (\ref{effIG}) is in agreement with the same obtained
in ref. \cite{Quan03}, where the quantum Otto cycle was considered
for a single particle into a one-dimensional box.

\subsection{The Isoenergetic Cycle for the GQW}

By analogy with other models which use an isoenergetic stroke to
build a quantum heat engine \cite{Pena01,Santos01,Pena02}, the isoenergetic
cycle based on the gravitational quantum well is depicted in fig.
\ref{Cycles}, and it is composed by two isoenergetic and two isoentropic
strokes. Again, it will be assumed that the isoenergetic strokes
are performed by changing the value of the gravitational field quasi-statically
in order to keep the expectation value of the Hamiltonian constant.
This requirement defines an energy bath. By considering the average
energy, $\left\langle E\right\rangle =\left\langle \hat{H}\right\rangle $,
one has,
\begin{equation}
\left\langle E\right\rangle =\sum_{n}p_{n}(g)E_{n}(g),\label{EEE}
\end{equation}
where we have written explicitly the dependence on the intensity
of $g$. The change in the average energy by considering quasi-static
strokes which depend exclusively on $g$ is given by \cite{Pena02},
\begin{eqnarray}
\delta\left\langle E\right\rangle  & = & \sum_{n}E_{n}(g)\frac{\partial}{\partial g}p_{n}(g)+\sum_{n}p_{n}(g)\frac{\partial}{\partial g}E_{n}(g),\nonumber \\
 & = & dQ+dW,\label{Quanti}
\end{eqnarray}
where the quantities
\begin{eqnarray}
\delta\left\langle Q\right\rangle  & = & E_{n}(g)\frac{\partial}{\partial g}p_{n}(g),\label{deltaQ}\\
\delta\left\langle W\right\rangle  & = & \sum_{n}p_{n}(g)\frac{\partial}{\partial g}E_{n}(g)\label{deltaW}
\end{eqnarray}
have been defined.

Here, it is important to note that, although eq. (\ref{Quanti}) reflects
the first law of thermodynamics, the quantity $\delta\left\langle Q\right\rangle $
is traditionally associated with a well defined temperature of the
system when in contact with a thermal reservoir. As this is not the
case for an isoenergetic stroke, where the system is in contact with
an energy reservoir, $\delta\left\langle Q\right\rangle $ is known
as the energy exchange or simply the heat exchange for convenience
of language, while $\delta\left\langle W\right\rangle $ is the work
done/performed by/on the system. 

By taking into account that work is done when $p_{n}(g)$ is kept
fixed, we can obtain explicitly an expression for the work as been,
\begin{eqnarray}
\left\langle W\right\rangle _{k\rightarrow\ell} & = & \int_{g_{k}}^{g_{\ell}}p_{n}(g)\frac{\partial E_{n}(g)}{\partial g}dg|_{p_{n}(g)\,\mbox{fixed}}\nonumber \\
 & = & p_{n}(g_{k})[E_{n}(g_{\ell})-E_{n}(g_{k})].\label{work}
\end{eqnarray}

For the isoenergetic stroke, the system-reservoir heat exchange
can be obtained analytically from (\ref{deltaQ}) and, by considering
that the system starts the cycle at the ground state with $p_{1}(g_{1})=1$,
and performs a maximal expansion and maximal compression, one has
\cite{Pena01,Santos01},
\begin{align}
\left\langle Q\right\rangle _{1\rightarrow2} & =E_{1}(g_{1})ln\left[\frac{E_{1}(g_{2})-E_{2}(g_{2})}{E_{1}(g_{1})-E_{2}(g_{1})}\right]\nonumber \\
 & +\int_{g_{1}}^{g_{2}}\frac{E_{1}\frac{dE_{2}}{dg}-E_{2}\frac{dE_{1}}{dg}}{E_{1}(g)-E_{2}(g)}dg,\label{Q in}
\end{align}
for maximal expansion and, 
\begin{align}
\left\langle Q\right\rangle _{3\rightarrow4} & =E_{2}(g_{3})ln\left[\frac{E_{2}(g_{4})-E_{1}(g_{4})}{E_{2}(g_{3})-E_{1}(g_{3})}\right]\nonumber \\
 & +\int_{g_{3}}^{g_{4}}\frac{E_{2}\frac{dE_{1}}{dg}-E_{1}\frac{dE_{2}}{dg}}{E_{2}(g)-E_{1}(g)}dg,\label{Q out}
\end{align}
for maximal compression.

With the analytical expressions for work and heat, let one describe
the isoenergetic cycle for the GQW in detail. The working substance
starts at the ground state $\psi_{1}(g_{1})$ and with energy $E_{1}(g_{1})$.
The first stroke is an isoenergetic expansion from $\ell_{g_{I}}$ to
$\ell_{g_{II}}$. Considering the maximal expansion and defining an
expansion coefficient, $c_{1}\equiv\ell_{g_{II}}/\ell_{g_{I}}$, the
isoenergetic stroke leads to,
\begin{equation}
E_{1}(g_{I})=E_{2}(g_{II}),\label{eq01}
\end{equation}
which results in,
\begin{equation}
c_{1}=\frac{a_{2}}{a_{1}}.\label{c1}
\end{equation}

For the eigenenergies given by (\ref{Ener02}), the second term on
the right-hand side in (\ref{Q in}) and (\ref{Q out}) vanishes and for the isoenergetic
expansion the heat exchange is given by,
\begin{equation}
\left\langle Q\right\rangle _{I\rightarrow II}=E_{1}(g_{I})ln\left[\frac{a_{1}}{a_{2}}\right].\label{Q12A}
\end{equation}

The next stroke is an isoentropic compression characterized by $p_{2}(g_{II})=p_{2}(g_{III})=1$.
Like in the isogravitational cycle, it will be convenient to define
a compression coefficient expansion $\alpha\equiv\ell_{g_{III}}/\ell_{g_{II}}.$
The work performed in this stroke can be easily obtained using (\ref{work}).
The third stroke is an isoenergetic compression from $\ell_{g_{III}}$
to $\ell_{g_{IV}}$. By defining a compression coefficient, $c_{3}\equiv\ell_{g_{IV}}/\ell_{g_{III}}$
, the isoenergetic condition implies in,
\begin{equation}
E_{2}(g_{III})=E_{1}(g_{VI}),\label{eq02}
\end{equation}
which results in,
\begin{equation}
c_{3}=\frac{a_{1}}{a_{2}}.\label{c3}
\end{equation}

By solving eq. (\ref{Q out}) for the conditions above one obtains,
\begin{equation}
\left\langle Q\right\rangle _{III\rightarrow IV}=E_{2}(g_{I}/(c_{1}\alpha)^{3})ln\left[\frac{a_{2}}{a_{1}}\right].\label{Q34A}
\end{equation}

To complete the isoenergetic cycle, an isoentropic stroke is performed
from $\ell_{g_{IV}}$to $\ell_{g_{I}}$, such that the work performed
here can be again obtained using (\ref{work}). The thermodynamic
efficiency of this cycle is given by,
\begin{equation}
\eta=1-\left|\frac{\left\langle Q\right\rangle _{III\rightarrow IV}}{\left\langle Q\right\rangle _{I\rightarrow II}}\right|=1-\frac{a_{1}}{a_{2}}\frac{1}{\alpha^{2}}.\label{efIE}
\end{equation}

The efficiency for the isogravitational and isoenergetic cycles
are depicted in fig. \ref{FigEf}. From eq. (\ref{efIE}), it can
be observed that the ratio $a_{1}/a_{\ell}$ is the lowest possible
when $\ell=2$, which means that one can, in principle, improve the
efficiency of the isoenergetic cycle by modeling the effective two
level system considering the first excited state and other states
with $\ell>2$, but is not possible it surpasses the unit, a physical
meaning that must be fulfilled. Another point concerning the isoenergetic
cycle is that $\alpha=\ell_{g_{III}}/\ell_{g_{II}}=(g_{II}/g_{III})^{1/3}$
can be arbitrarily large, resulting this way in a higher value of
$\eta$. However, it is important to stress that the real value of
the efficiency is limited by the length scale of the system, i. e.
$\ell_{0}$. 

\subsection{Experimental Feasibility of the isogravitational and isoenergetic Cycles}

Once the isogravitational and the isoenergetic cycles for the GQW architecture have been described, it is important to discuss the potential feasibility of these models. Starting with the isoenergetic model, from an experimental standpoint, even being considered a trend topic affected by a plethora of quantum phenomena as entanglement, decoherence, etc, one suggests \cite{Beretta, Scully01} that the use of a maser-laser apparatus to implement the heat and work interaction simultaneously -- through a smooth continuous change of the gravitational field (or analogously, an electric field intensity), as it is performed to realize the isotherms of the Carnot-like cycles -- sheds some light on the possibility of engendering a similar mechanism which accounts for a fine-tuning between temperature and gravitational field intensity, as to generate an isoenergetic stroke. 

\begin{figure}
\includegraphics[scale=0.7]{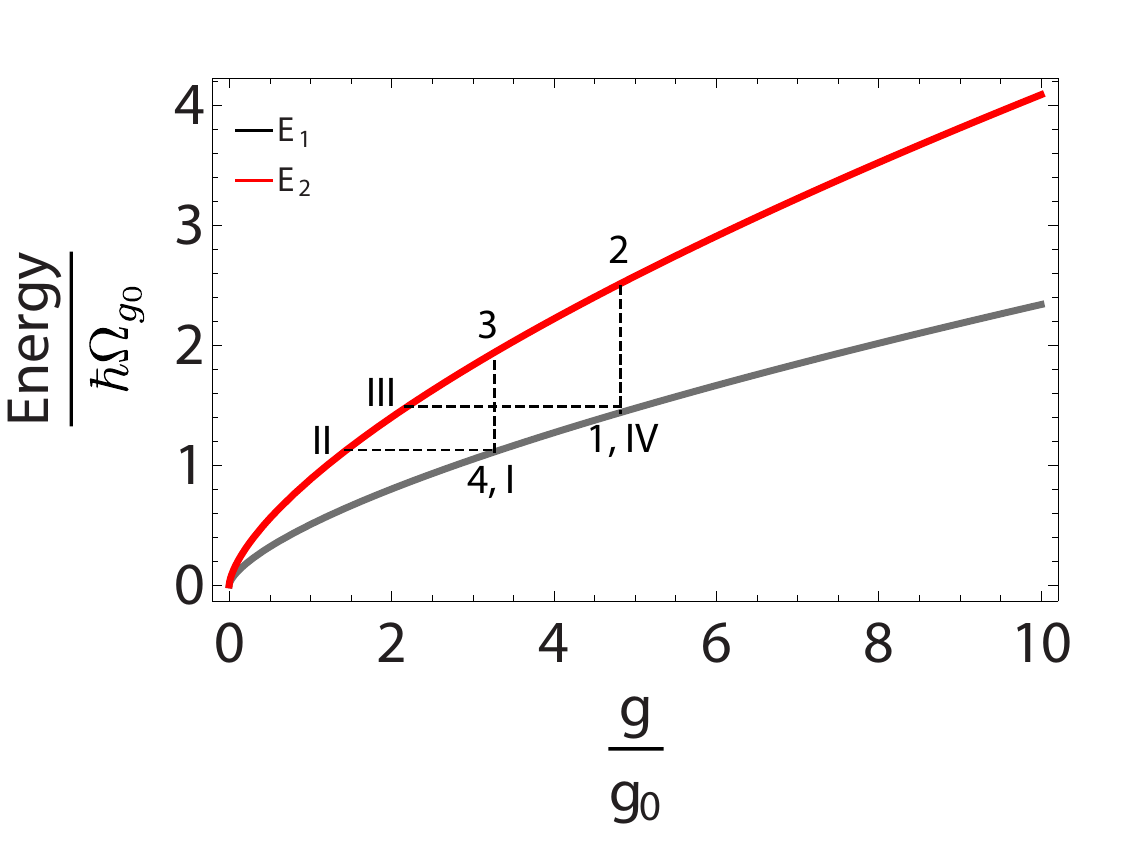}\caption{Isogravitational and isoenergetic cycles for the
effective two level system modeled by the ground and first excited
states of the GQW system. The isogravitational cycle involves two
isogravitational and two isoentropic strokes (Latin numbers) and
the isoenergetic cycles is composed by two isoenergetic and two
isoentropic strokes (Arabic numbers). It was considered $g_{0}=10m/s^{2}$.
\label{Cycles}}
\end{figure}

\begin{figure}
\includegraphics[scale=0.5]{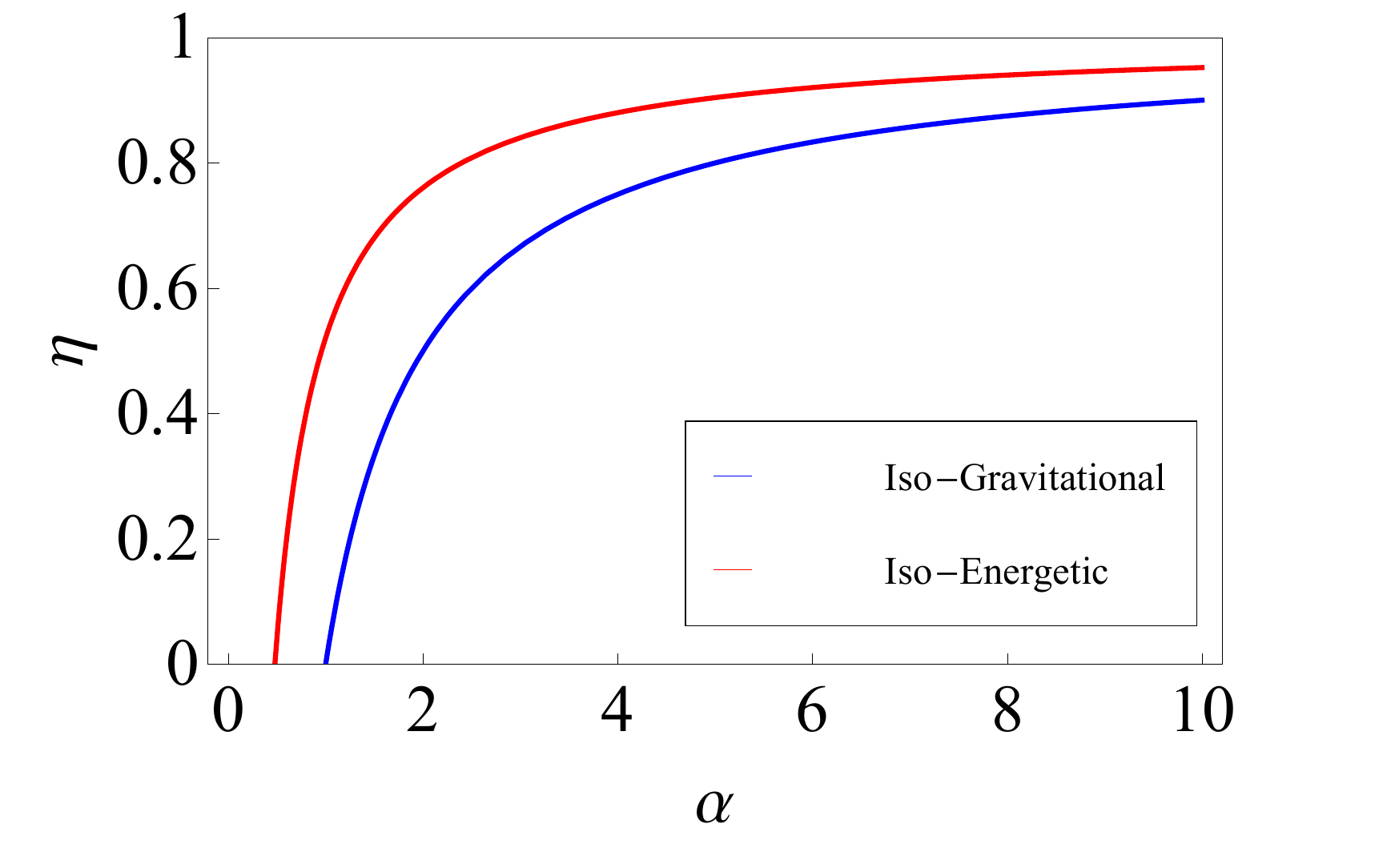}

\caption{Thermodynamic efficiency for the isogravitational
(red) and isoenergetic (blue) cycles for the gravitational quantum well
as an effective two level system.}
\label{FigEf}
\end{figure}

For the isogravitational cycle, the isoenergy gap cycle fueled by a particle trapped in a gravitational field, it is clear that one must include two thermal reservoirs into the engine, one to perform the hot isochore-like trajectory from $1$ to $2$ and another to perform the cold isochore-like trajectory from $3$ to $4$ as depicted in fig. \ref{Cycles}. By focusing on the entropy generation due to the processes concerning the isogravitational cycle, according to ref. \cite{Beretta}, the entropy balance reads:
\begin{equation}
dS =  \frac{\delta\overleftarrow{Q}}{T_h} - \delta S_{gen},
\end{equation}
where $\overleftarrow{Q}$ is the heat absorbed by the system and $T_h$ is the temperature of the hot bath. Considering the terms on r. h. s. of the above equation, the first corresponds to the increasing of entropy due to the heat interaction and the second one to the increasing entropy to the internal dynamics, for instance, relaxation and decoherence processes. Now, identifying the work performed by the working substance just as a function of the energy gap, the expression for $\delta S_{gen}$ is computed to be given by \cite{Beretta},
\begin{equation}
\delta S_{gen} =  \frac{\delta \overleftarrow{Q}}{T} - \frac{\delta \overleftarrow{Q}}{T_h},
\label{ddd}
\end{equation}
where $T$ is the temperature of the working substance and one notices that for an isoentropic stroke (from $2$ to $3$ and from $4$ to $1$ in fig. \ref{Cycles}) $\delta S_{gen} = 0$. 

For the isoenergy gap fueled by the GQW system as working substance, which the isogravitational stroke is engendered by thermal contacts with reservoir at temperatures $T_{high}$ and $T_{cold}$, one notices from eq. (\ref{ddd}) that it encompasses the entropy generation due to internal dynamics, namely due to an irreversible process. Nevertheless, it could be recovered, in principle, by a sequence of infinitesimal contacts with a infinite number of hot and cold bath covering the temperature of the working substance and the temperatures of the reservoirs, $T_{high}$ and $T_{cold}$, as mentioned in ref. \cite{Beretta}. This should assure a theoretical formulation of our isogravitational cycle. Another possibility of implementing experimentally the isoenergy gap fueled by the GQW system is well illustrated in ref. \cite{Nori}, where is showed that a quantum Otto cycle can be modeled as an infinity number of quantum Carnot cycle, with different temperatures of the two bath of these quantum Carnot cycles. In our notation, the isogravitational stroke (from $1$ to $2$ and from $3$ to $4$ in fig. \ref{Cycles}) can be modeled as many small quantum isothermal strokes.

\section{general isoenergetic cycle for any pair of eigenstates\label{sec:general-iso-energetic-cycle}}

\begin{figure}
\includegraphics[scale=0.5]{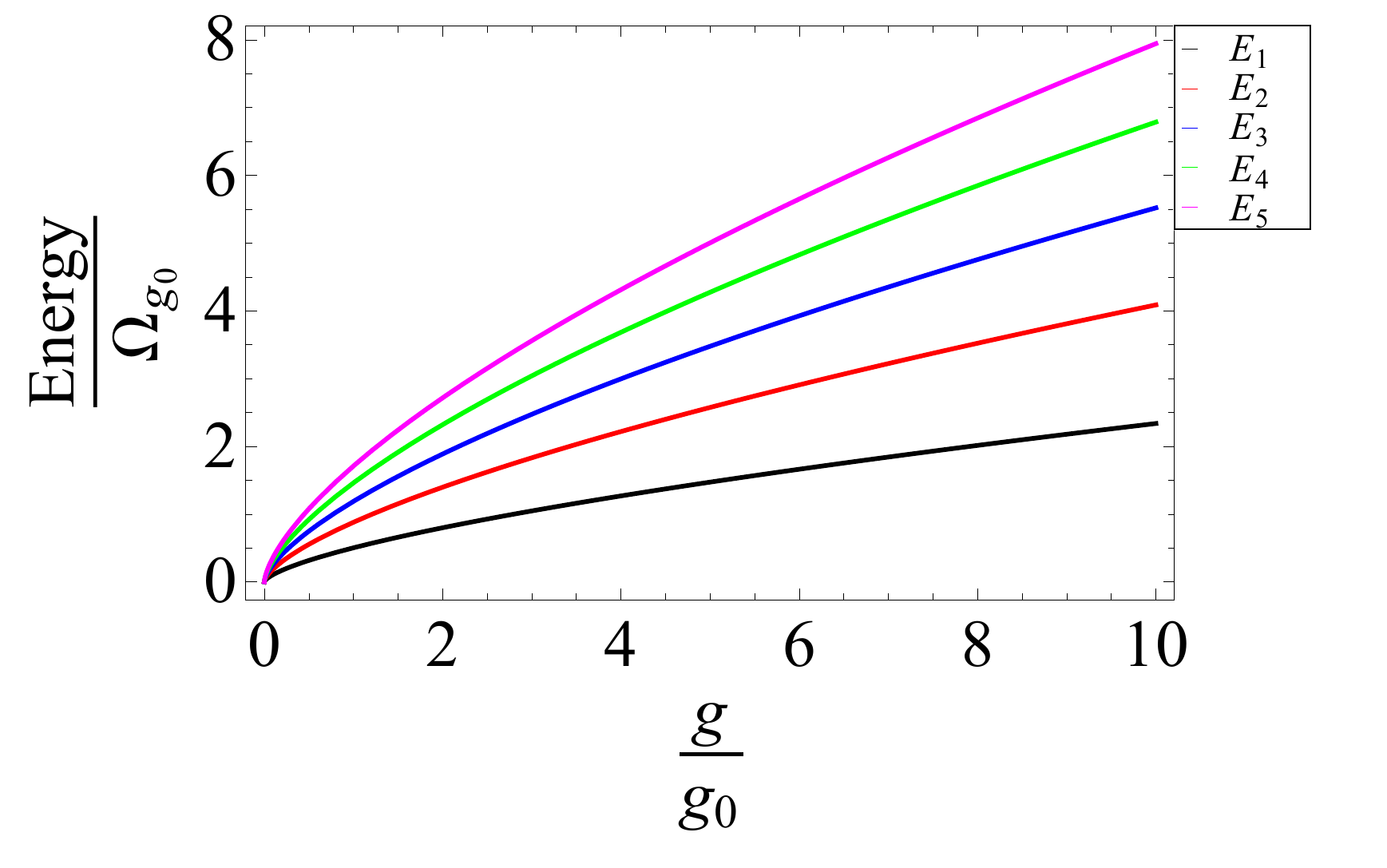}

\caption{First five eigenenergies of the GQW system to illustrate
the possibility of generating an effective two level system with any
pair of eigenstates. The cycle can be performed with any two arbitrary
states. It was considered $g_{0}=10m/s^{2}$.}

\label{Fig11}
\end{figure}

Here we will generalize our analysis of section $III$ for the case
of arbitrary pair of eigenstates. The idea is to show that it is possible,
in principle, to model an isoenergetic cycle for the GQW system with
any pair of eigenstates $\psi_{k}(g_{i})$ and $\psi_{\ell}(g_{j})$
and thus obtain a general relation for the efficiency. From eqs. (\ref{Q12})
and (\ref{Q34}), the heat exchange between the system and the energy
reservoirs for general eigenstates are given by,
\begin{eqnarray}
\left\langle Q\right\rangle _{n\rightarrow m} & = & E_{n}(g_{a})\mbox{ln}\left[\frac{E_{n}(g_{b})-E_{m}(g_{b})}{E_{n}(g_{a})-E_{m}(g_{a})}\right],\label{Qnm}\\
\left\langle Q\right\rangle _{m\rightarrow n} & = & E_{m}(g_{c})\mbox{ln}\left[\frac{E_{m}(g_{d})-E_{n}(g_{d})}{E_{m}(g_{c})-E_{n}(g_{c})}\right],\label{Qmn}
\end{eqnarray}
where a general notation was introduced, $g_{b}<g_{a}$
and $g_{c}<g_{d}$, with $E_{n}$ and $E_{m}$ being the corresponding
energies to the first and second arbitrary eingenstates, respectively.
Note that the second term on right in (\ref{Q in}) and (\ref{Q out})
vanishes independent of the choice of the eigenstates. The fig. \ref{Fig11}
illustrates the first five eigenenergies of the GQW system in order
to show the possibility of generating an effective two level system
considering any pair of eigenstates.

By solving these expressions using the eigenenergies (\ref{Ener02}),
one has,
\begin{eqnarray}
\left\langle Q\right\rangle _{n\rightarrow m} & = & E_{n}(g_{a})\mbox{\mbox{ln}}\left[a_{n}/a_{m}\right],\label{Qnm01}\\
\left\langle Q\right\rangle _{m\rightarrow n} & = & E_{m}(g_{c})\mbox{\mbox{ln}}\left[a_{m}/a_{n}\right],\label{Qmn01}
\end{eqnarray}
with $a_{n(m)}$ being the root of the Airy function in (\ref{eigenfu03}).
Thus, through the thermodynamic efficiency, one obtains a general
expression for an isoenergetic cycle for the GQW system which is
valid for any pair of eigenstates and is given by,
\begin{equation}
\eta_{General}=1-\frac{E_{m}(g_{c})}{E_{n}(g_{a})}.\label{General}
\end{equation}

The equation (\ref{General}) is general and is in complete agreement
with the analogous result obtained in ref. \cite{Bender03}, which
introduces the notion of energy reservoir. We stress that this equation
is equivalent to the Carnot efficiency in the sense that the energy
assumes the role of the temperature and thus it is analogous to the
classical thermodynamic result. Note that we find this result without
any assumptions on the expansion coefficient during the process. The
only requirement is that the strokes are performed quasi-statically.

\section{Equation of State for GQW Isoenergetic Cycle\label{sec:Equation-of-State}}

Now, we want to describe the isoenergetic stroke for the GQW system
through an equation of state. First of all, using the length scale
for the gravitational well, $\ell_{g}=(\hbar^{2}/2m^{2}g)^{1/3}$,
the eigenenergies can be written as follows,
\begin{equation}
E_{n}(g)=-\frac{\hbar^{2}}{2m\ell_{g}^{2}}\alpha_{n}.\label{Ell}
\end{equation}

During the first stroke, the isoenergetic expansion, the work done
can be attributed to a force-like external agent on the wall of the
gravitational well. The contribution to this force from the nth energy
eigenstate is defined as,
\begin{equation}
f_{n}=-\frac{\hbar^{2}}{m\ell_{g}^{3}}\alpha_{n},\label{fn}
\end{equation}
such that the force $F$ is given by the expectation value,
\begin{equation}
F=\sum_{n}p_{n}f_{n}.\label{P}
\end{equation}

Using the eq. (\ref{EEE}) for the expectation value of energy and
considering that the working substance starts the cycle at $E_{1}(g_{a})$,
one has that $\sum_{n}p_{n}E_{n}(g_{a})=E_{1}(g_{a})$, thus we get,
\begin{equation}
F\:\ell_{g}=2E_{1}(g_{a}),\label{statee}
\end{equation}
where, multiplying both sides for $\ell_{g}^{2}$, one has,
\begin{equation}
P\:\ell_{g}^{3}=2E_{1}(g_{a}),\label{stateee}
\end{equation}
having defined $P=F/\ell_{g}^{2}$ as an analogous of pressure and
$\ell_{g}^{3}$ as an equivalent of volume. Equation (\ref{stateee})
is similar to the corresponding equation of state for an isothermal
process of a classical ideal gas when the analogies $2E_{1}(g_{a})\leftrightarrow k_{B}T$
and $\ell_{g}^{3}\leftrightarrow V$ are performed. This result reinforces
the conceptual definition of thermal bath summarized in the eq. (\ref{General})
which replaces the temperatures by the eigenenergies in the efficiency.

\section{Conclusions \label{sec:Conclusions}}

In this work the possibility of modeling the gravitational
quantum well as an effective two level system (TLS) was explored. By considering two
cycles for the TLS, the isogravitational and the isoenergetic ones,
it was possible to verify analytically how the efficiency
behaves when varying the isoentropic expansion coefficient, $\alpha$,
defined in terms of a length scale for the GQW system. The natural
length for the system is $\ell_{0}=(\hbar^{2}/2m^{2}g_{0})^{1/3}$,
which is approximately $5,87\mu m$ if we consider the mass of the
neutron and the gravitational acceleration of the Earth, i. e.
greater than the typical size of a semiconductor quantum dot \cite{dot},
$\ell_{d}=70nm$, evidencing, in principle, the physical realization
of the model presented here. Another motivation to perform experimentally
the isogravitational and the isoenergetic cycles, is given in ref. \cite{Batalhao}, where
a high level of control was demonstrated in implementing a quantum
heat engine based on a spin-$1/2$ system in a nuclear magnetic resonance
apparatus. However, due to the extremely weak intensity of the gravitational
field, the efficiency shall be very small. In this point, it is important
to mention that the general scheme developed here is useful also for the electric field, which can be conveniently controlled
in the laboratory in terms of intensity and the particle confined
into the potential could be an electron.

A general expression for an isoenergetic cycle engendered
with any pair of eingenstates of the gravitational quantum well system was also obtained
and it was shown that the equation is in concordance with the obtained
in ref. \cite{Bender03}, where the concept of energy bath was introduced.
This result shows that the efficiency of the isoenergetic cycle for
the GQW system will depend only on the ratio of the two roots of the
Airy function and can be modified depending on the distance between
the two chosen eigenstates.

As a final result, a relevant relation for the isoenergetic
stroke was derived, an analogous equation of state for this process, which is
similar to the equation for the isothermal process of an ideal gas.
This relation reinforces the substitution of the temperature of the
heat bath by the energy concept in the energy bath. Finally, we believe
that this application of the gravitational quantum well as a fuel
for quantum heat engines could be useful to encourage the application
of quantum thermodynamics concepts to this system such as the possibility
of generating entangled states or non-equilibrium strokes.

\begin{acknowledgments}
JFGS would like to thank CAPES (Brazil) and Federal University of
ABC (UFABC) for support. The author also thanks the PhD Patrice A. Camati
from UFABC for relevant discussion and advice.
\end{acknowledgments}

\newpage

\end{document}